\documentclass[twocolumn,pra,preprintnumbers,amsmath,amssymb]{revtex4}
\newcommand{\p}{\partial}

\begin{document}

\preprint{RIKEN-TH-156, DAMTP-2009-44}

\title{Gravitational Dual of Tachyon Condensation}

\author{Gary W. {\sc Gibbons}}\email[]{G.W.Gibbons(at)damtp.cam.ac.uk}
\affiliation{
{\it DAMTP, University of Cambridge,
Wilberforce Rd, Cambridge CB30WA, U.K.}}

\author{Koji {\sc Hashimoto}}\email[]{koji(at)riken.jp}
\affiliation{
{\it Theoretical Physics Laboratory, Nishina Center, RIKEN, Saitama 351-0198,
Japan }}

\author{Shinji {\sc Hirano}}\email[]{hirano(at)nbi.dk}
\affiliation{{\it The Niels Bohr Institute, Blegdamsvej 17, DK-2100,
Copenhagen, Denmark}}

\begin{abstract}
We study a system of $N$ D3-branes in which open string tachyons survive in the low energy $\alpha'\to 0$ limit. We compute the height of the tachyon potential both in the weak and strong  couplings by using ${\cal N}= 4$ super Yang-Mills and the dual AdS$_5$ descriptions respectively. We find an exact agreement between the two descriptions in the large $N$ limit. This provides an example of gravitational duals of open string tachyon condensation.
\end{abstract}

\maketitle

\section{Introduction}

\noindent
There has been considerable progress in understanding non-perturbative aspects of string theory with or without supersymmetries.  These developments culminated in the gauge/string duality primarily in supersymmetric cases \cite{Maldacena:1997re}. Meanwhile, in non-supersymmetric cases one of the most remarkable progress is the open string tachyon condensation \cite{senconj}.
However, these two developments have been rather orthogonal to each other, except for two-dimensional string theories \cite{McGreevy:2003kb}.
In this note we study an open string tachyon condensation in the gauge/string duality and provide an example in which these two connect at the quantitative level \footnote{A discussion on unstable D-branes in AdS/CFT previously appeared in \cite{Drukker:2000wx}.}.

In formulating tachyon condensations in the gauge/string duality, one immediate obstacle is that the mass squared of open string tachyons is of order ${\cal O}(1/\alpha')$ which may not be visible in the field theory limit $\alpha'\rightarrow 0$.
However, there exist systems in which the open string tachyon mass remains of order ${\cal O}(1)$ (in an appropriate unit) in the limit $\alpha'\to 0$.
One such example is a system of intersecting branes at angles. The mass squared of open string tachyons stretched between intersecting branes is proportional to $\theta/\alpha'$ where $\theta$ is an angle. 
In this case we can take the $\alpha'\rightarrow 0$ limit, keeping $\theta/\alpha'$ fixed finite.
This type of open string tachyons manifest themselves as unstable modes in the gauge theory in the low energy $\alpha'\to0$ limit of intersecting D-branes \cite{Hashimoto:1997gm}. 
In the present paper we consider a T-dual version of the system studied in \cite{Hashimoto:1997gm} in the 't Hooft limit, $N\to \infty$ and $g_{YM}\to 0$ with $\lambda=g_{YM}^2N$ fixed finite. 

One of the most important aspects of open string tachyon condensation is that unstable D-branes which support open string tachyons disappear after the tachyons are condensed.
Accordingly, Sen's conjecture states that the height of the tachyon potential equals the tension (energy density) of unstable D-branes \cite{senconj}.
Hence the height of the tachyon potential is one of the most important and unambiguous quantities in the open string tachyon condensation. 
The main objective of this paper is to compute the height of the tachyon potential both in the gauge theory and the dual AdS gravity for a large $N$ D3-brane configuration and to see if they agree.  
We indeed find an exact agreement in the large $N$ limit, thus providing a simple example of gravitational duals of open string tachyon condensation.

\section{Brane configuration}

\noindent
We consider a brane configuration of a D3-D1 bound state parallel to coincident 
$N$ D3-branes at some distance.
The low energy description of this system is given by ${\cal N}=4$ $U(N+1)$ super Yang-Mills (SYM) theory. A D1-brane on D3-branes is represented by a magnetic flux.
Our brane configuration is thus described by the following gauge and adjoint scalar fields
\begin{equation}
\begin{array}{l}
F_{23} = {\rm diag}(B,0,0,\cdots) \ , \\
\Phi  =  {\rm diag}(\phi,0,0,\cdots) \ , 
\end{array}
\label{initial}
 \end{equation}
where the magnetic flux $B$ is T-dual to the aforementioned angle $\theta$, and $B$ and $\phi$ are the constant parameters.
This is a classical solution of the SYM theory. It should be clear that the D3-D1 bound state is separated from the $N$ D3-branes by the distance $2\pi\alpha' \phi$.

When the D3-D1 bound state is close to the $N$ D3-branes ($\sqrt{\alpha'}\phi\ll 1$), there
appear tachyonic fluctuation modes in the off-diagonal 
entries of the gauge and scalar fields. 
This is a manifestation of string excitations connecting the D3-D1 bound state and the $N$ D3-branes \cite{Hashimoto:1997gm} (see also  \cite{Hashimoto:2003xz} for its relevance to D-brane reconnections). 
More precisely, the lowest excitation of these strings has the mass squared
\begin{eqnarray}
 m^2= \left(2\pi\alpha' \phi {\cal T}\right)^2 - 
\frac{\arctan(2\pi \alpha' |B|)}{2\pi\alpha'}\ ,
\end{eqnarray}
where ${\cal T}=1/2\pi\alpha'$ is the string tension. For small 
$|B| \ll 1/\alpha'$, this yields
\begin{eqnarray}
 m^2 = \phi^2 - |B|.
\label{tacmass}
\end{eqnarray}
Thus a tachyon develops for $\phi < \sqrt{B}$.
We would like to stress that this tachyon has survived in the field theory limit $\alpha' \rightarrow 0$.
Hence it is visible in the gauge/string duality where the low energy $\alpha'\to 0$ limit of $N$ D3-branes is dual to the string theory on $AdS_5\times S^5$.
Note that when $B=0$, this configuration is a point in the Coulomb branch of
${\cal N}=4$ SYM. When $B\ne 0$, the Coulomb branch is
lifted and the configuration is no longer supersymmetric.

\section{The potential height at weak coupling -- field theory}

\noindent
We now compute the height of the tachyon potential based on the Sen's conjecture.
At weak coupling $\lambda=g_{YM}^2N\ll 1$ we can use the classical YM theory to evaluate
the potential height. The top of the tachyon potential corresponds to the $\phi\to\infty$ limit at which the D3-D1 bound state and the $N$ D3-branes are infinitely separated and do not feel an attractive force 
(in terms of ${\cal N}=4$ SYM theory, the one-loop potential vanishes for 
$\phi \rightarrow \infty$). The total energy density is given by
\begin{eqnarray}
 {\cal E}\bigm|_{\phi=\infty}
={N B^2\over 2\lambda}\ .
\end{eqnarray}
When $\phi<\sqrt{B}$, the tachyon develops. 
The endpoint of the tachyon condensation is a bound state of $N+1$ D3-branes and D1-branes
 (this is a T-dual of D-brane reconnections in \cite{Hashimoto:2003xz}).
Initially the D1-branes are bound to a single D3-brane isolated from the $N$ coincident D3-branes.
The D3-D1 bound state is attracted to the $N$ D3-branes ($\phi$ decreases) and they coalesce ($\phi=0$). In this process the D1-branes distribute themselves uniformly among $N+1$ coincident D3-branes.
The total D1-brane charge  ${\rm tr} F_{23}$ is conserved in the process. Thus the D1-brane charge per each D3-brane is $B/(N+1)$; 
\begin{equation}
\begin{array}{l}
F_{23} = {\rm diag}\left({B\over N+1},{B\over N+1},\cdots, {B\over N+1}\right) \ , \\
\Phi =  {\rm diag}(0,0,\cdots, 0) \ . 
\end{array} 
\end{equation}
Hence the energy density of the final state is 
\begin{eqnarray}
{\cal E}\bigm|_{\phi=0} 
={N+1\over 2g_{YM}^2}\left({B\over N+1}\right)^2
={NB^2\over 2\lambda}{1\over N+1}\ .
\end{eqnarray}
At large $N$, the difference of the energy yields
\begin{eqnarray}
  {\cal E}\bigm|_{\phi=\infty} 
- {\cal E}\bigm|_{\phi=0} 
={NB^2\over 2\lambda}+{\cal O}(1)\ .
\label{height}
\end{eqnarray}
This is the height of the tachyon potential at weak coupling.
Note that since the effective coupling constant of the AdS string is $1/N$, the $N$ dependence is as expected for the D-brane tension.  

Note also that we could have computed the potential height without introducing $\phi$.
However, the role of $\phi$ is to dial the instability of our unstable brane system and catalyse the process of tachyon condensation.
In the dual gravity description, as usual, the energy scale $\phi$ is mapped to the radial distance.

\section{The potential height at strong coupling -- gravity dual}

\noindent
We next compute the potential height at strong coupling $\lambda=g_{YM}^2N=2\pi g_sN\gg 1$. 
The low energy limit of $N$ D3-branes is dual to the string theory on $AdS_5\times S^5$.
So in the dual description at $\lambda\gg 1$ we treat the D3-D1 bound state as a probe D3-brane with a magnetic flux put in the $AdS_5\times S^5$ geometry. 

The low energy effective action of a probe D3-brane in the 
${\rm AdS}_5\times {\rm S}^5$ geometry is 
given by the Dirac-Born-Infeld action plus the
Chern-Simons term (Ramond-Ramond 4-form coupling) with the ansatz, $r=r(t,{\bf x})$.
The AdS$_5$ metric  is 
\begin{eqnarray}
ds^2 =\frac{r^2}{R^2} ( -dt ^2 + d {\bf x} ^2)  + \frac{R^2}{r^2}dr^2\ ,
\end{eqnarray}
where the radius $R =  (4\pi g_s (\alpha')^2 N)^{1/4}$ and $r =0$ is the horizon. The resultant action is
\begin{equation}
S = -{\cal T}_{D3} \int d^4x\,
f^{-1}\left[
\sqrt{-\det H_{\alpha\beta}}-1
\right]\ ,
\end{equation}
where
\begin{eqnarray}
H_{\alpha\beta}&=&\eta_{\alpha\beta}+ f \p_\alpha r \p_\beta r
+ 2\pi\alpha' \sqrt{f}F_{\alpha\beta}\ ,\\
 f &=& \frac{R^4}{r^4} \ .
\end{eqnarray}
In the decoupling limit the coordinate $U\equiv r/\alpha'$ is fixed finite, while $\alpha'$ is taken to zero.
We consider the constant $U$ and the magnetic field $F_{23}=B$.
In these variables the action takes the form
\begin{eqnarray}
 S = \frac{-1}{(2\pi)^3g_{\rm s}}
\int \! d^4x \;
\frac{U^4}{2\lambda}\!
\left[
\sqrt{1 + \frac{2(2\pi)^2 \lambda B^2}{U^4}}-1
\right] \ .
\end{eqnarray}
As expected, this shows that the D3-D1 bound state falls into the AdS horizon.
As mentioned above, the energy scale $\phi$ in the gauge theory corresponds to the radial coordinate $U$ in the gravity. So we can read off the potential energy from this expression. 
In the large $U$ (thus large $\phi$) limit, the energy density yields
\begin{eqnarray}
{\cal E}\bigm|_{U\gg 1}  = \frac{1}{4\pi g_{\rm s}}
\left[B^2 + {\cal O}(1/U^4)\right]\ .
\end{eqnarray}
In the small $U$ (thus small $\phi$) limit, it yields
\begin{equation}
{\cal E}\bigm|_{U\ll 1} = \frac{1}{(2\pi)^3g_{\rm s}}
\left[2\pi \frac{|B|}{\sqrt{2\lambda}}U^2 + {\cal O}(U^4)
\right]\ .
\end{equation}
Therefore, the height of the potential is given by
\begin{eqnarray}
{\cal E}\bigm|_{U=\infty}-{\cal E}\bigm|_{U= 0} = \frac{B^2}{4\pi g_{\rm s}}
={NB^2\over 2\lambda}\ . 
\end{eqnarray}
We thus find an exact agreement with the weak coupling field theory result (\ref{height}) in the large $N$ limit. 
Note, however, that the value of the potential at the bottom differs -- $B^2/2\lambda$ in the field theory  and zero in the gravity.

As we have just seen, the AdS description of the (lifted) Coulomb branch of ${\cal N}=4$ SYM  encodes the height of the tachyon potential.
However, the open string tachyon is only a valid picture in the weak coupling field theory, as it appears as a small fluctuation mode of the gauge and scalar fields.
So in the strong coupling one cannot directly observe the open string tachyon.
In other words, the gauge/gravity duality is an open string (loops)/closed string (tree) channel duality. 
Hence in the gravity description the open string tachyon is integrated out and cannot be directly observed \footnote{Some discussion on the absence of tachyon modes in the Coulomb branch
can be found in \cite{Silverstein:2003hf}.}.
However, there is no contradiction in the agreement we found. The height of the tachyon potential is the energy of an unstable brane and conserved at any gauge couplings.
So the potential height is an observable in the gravity description.

It is, however, noteworthy that the value of the potential height in the
large $N$ limit is not renormalized, remaining the same value from weak
to strong coupling  \footnote{A similar agreement was found in a
different duality frame \cite{Hashimoto:2000ys}. Our brane configuration
is T-dual to that, and also to \cite{Takahashi:2004np}.}.
This calls for an explanation. The reason may be that the magnetic flux $B$ on the D3-brane is small $B\ll 1/\alpha'$ in the D-brane set-up prior to the decoupling (field theory) limit. Hence our system of a D3-D1 bound state parallel to $N$ D3-branes at some distance is close to a supersymmetric configuration.
It is often the case that not only BPS states but also near-BPS states are protected against stringy $\alpha'$ corrections \cite{Douglas:1996yp}. In the decoupling limit the effective $\alpha'_{eff}$ of the AdS string is $1/\sqrt{\lambda}$.
We thus believe that this is the reason why we find an exact agreement between the gauge theory and the gravity results in the large $N$ limit.

\section{Summary and discussions}

\noindent
In this note we provide an example of open string tachyon condensations in the gauge/string duality.
The idea is to consider a brane configuration in which open string tachyons survive in the field theory $\alpha'\to 0$ limit.
One can explore various similar brane set-ups by generalising the example we presented. However, we find it more illuminating only to discuss a single simple example.

We would like to give a brief discussion on topological defects in the tachyon condensation and make a comparison between our study and the geometric tachyon observed by Kutasov \cite{Kutasov:2004ct}.

First we discuss topological defects in the tachyon condensation \cite{senconj}.
In the present set-up the topological defect is a D(-1)-brane. 
The reason is as follows: 
Our brane configuration consists of a D3-D1 bound state and $N$ D3-branes. 
The tachyonic modes come from 1-3 and 3-1 strings stretched between the D1-brane dissolved in the D3-D1 bound state and the $N$ D3-branes. So the tachyon in question is a complex tachyon. When the complex tachyon condenses, the topological defect is co-dimension 2 relative to the parent unstable brane.
In ${\cal N}=4$ SYM the tachyonic modes about the background (\ref{initial}) are the ground state of the Landau level and localized in the $(x^2,x^3)$-plane orthogonal to the magnetic flux $B$. Hence, the parent unstable brane is effectively a D1-brane extended along the $(x^0,x^1)$-directions.  
We thus expect a D(-1)-brane descended from the tachyon condensation as a vortex on the D1-brane. 
In the weak coupling it is possible to analyse the vortex solution. 
However, it seems rather difficult to study the nucleation of this topological defect in the gravity dual, since we cannot directly see the tachyon.
In general it would be challenging but very interesting to study the nucleation of topological defects in the gravity dual and compare them with the field theory counterparts.

Finally we make a brief comment on the geometric tachyon. 
Kutasov made an intriguing observation in \cite{Kutasov:2004ct} that the D$p$-brane effective action in the $k=2$ NS5-branes throat on a transverse circle coincides with the effective action of the non-BPS D$(p+1)$-brane \cite{Sen:1999md} wrapped on the circle in the zero radius limit. 
The transverse circle scalar on the D$p$-brane is identified with the tachyon on the non-BPS D$(p+1)$-brane. It was further argued that the two descriptions become identical via a second order phase transition \cite{Sen:2007cz}.
In comparison, in our case the geometric (radion) instability in the AdS space is related to the open string tachyon via the gauge/string duality. However, it is not possible to map directly between a geometric (radion) mode on one side and a tachyonic mode on the other, as discussed above.
Hence the geometric manifestation of tachyon condensation in our study is rather different from the geometric tachyon and they are not related to each other in any obvious ways.

\section*{Acknowledgment}

\noindent
K.H.~would like to thank DAMTP, University of Cambridge, for a kind
hospitality. K.H.~appreciate comments by S.~Terashima and S.~Sugimoto. 
K.H.~is partly supported by
the Japan Ministry of Education, Culture, Sports, Science and
Technology.
The work of S.H. was in part supported by FNU via grant number
272-06-0434.

\newcommand{\J}[4]{{\it #1} {\bf #2} (#3) #4}
\newcommand{\andJ}[3]{{\bf #1} (#2) #3}
\newcommand{\AP}{Ann.\ Phys.\ (N.Y.)}
\newcommand{\MPL}{Mod.\ Phys.\ Lett.}
\newcommand{\NP}{Nucl.\ Phys.}
\newcommand{\PL}{Phys.\ Lett.}
\newcommand{\PR}{ Phys.\ Rev.}
\newcommand{\PRL}{Phys.\ Rev.\ Lett.}
\newcommand{\PTP}{Prog.\ Theor.\ Phys.}
\newcommand{\hep}[1]{{\tt hep-th/{#1}}}

\end{document}